# The aDORe Federation Architecture

Herbert Van de Sompel [1], Ryan Chute [1], Patrick Hochstenbach [2]

[1] *Digital Library Research and Prototyping Team*

*Los Alamos National Laboratory*

*MS P362, PO Box 1663*

*Los Alamos, NM 87544-7113, USA*

{herbertv,rchute}@lanl.gov

[2] *Universiteitsbibliotheek*

*Universiteit Gent*

*Rozier 9, 9000 Gent, Belgium*

patrick.hochstenbach@ugent.be

Abstract

The need to federate repositories emerges in two distinctive scenarios. In one scenario, scalability-related problems in the operation of a repository reach a point beyond which continued service requires parallelization and hence federation of the repository infrastructure. In the other scenario, multiple distributed repositories manage collections of interest to certain communities or applications, and federation is an approach to present a unified perspective across these repositories. The high-level, 3-Tier aDORe federation architecture can be used as a guideline to federate repositories in both cases. This paper describes the architecture, consisting of core interfaces for federated repositories in Tier-1, two shared infrastructure components in Tier-2, and a single-point of access to the federation in Tier-3. The paper also illustrates two large-scale deployments of the aDORe federation architecture: the aDORe Archive repository (over 100,000,000 digital objects) at the Los Alamos National Laboratory and the Ghent University Image Repository federation (multiple terabytes of image files).

# Introduction

There is a growing interest in issues of scalability that are faced when designing, deploying, and managing infrastructures for ingesting, storing, accessing, and providing services for collections of digital objects. This increased interest in scalability is directly related to the exponential growth in the amount of digital artifacts that is being created on a daily basis, both born-digital, and as a result of massive digitization efforts. Architects, engineers and developers involved in creating digital asset management systems are facing the harsh reality that their solutions need to handle an amount of artifacts that is orders of magnitude higher than originally intended, and are reaching an understanding that approaches that work at the originally intended scale do not necessarily work at that next level. Whereas scalability used to be a concern for a limited group of traditional custodians of vast content collections, it is rapidly appearing on the radar of a much larger group of institutions worldwide, for example, as a result of their involvement in digitization projects, eScience and eHumanities data curation activities, digital preservation endeavors, and institutional repository efforts.

Scalability in digital libraries is a problem that extends into multiple dimensions. For example, there are issues related to the amount of digital objects to be handled and issues related to their size. There are issues related to the performance of processes such as ingestion of objects into a repository, dissemination of stored objects, and introspection upon stored objects among others driven by preservation requirements. Optimizing, tuning, and tweaking the existing repository infrastructure can initially alleviate performance problems, but eventually limits are reached. At that point, a major redesign of the repository solution is an obvious option. An alternative is to move towards an environment that consists of parallel instances of the existing repository solution and to glue those together into a repository federation that behaves as if it were a single repository. The desire to federate repositories in such a manner actually also emerges as a result of the understanding that no single digital library hosts all artifacts that are relevant for a specific subject domain, community, or application. The proposition of a "single repository behavior" exposed by a federation consisting of any number of distributed repositories is appealing, and has been the subject of digital library interoperability efforts such as Dienst [22], NCSTRL [8], CORDRA [36, 33, 15], DRIVER [9], and the Chinese DSpace federation [38]. Both federation paths, on one hand the federation of multiple instances of a specific repository installation, and on the other hand the federation of distributed repositories, reveal another dimension of the scalability problem in contemporary digital library efforts. Indeed, as a result of a combination of low-level system scalability issues, and higher-level community needs, there comes a point at which the reality of a multiple-repository environment must be embraced. The challenge is then to devise an approach to federate repositories in a manner that is functional, practically achievable, and … scaleable to a vast amount of federated repositories.

This paper describes the aDORe repository federation architecture, an outcome of the aDORe research and development effort by the Digital Library Research & Prototyping Team of the Los Alamos National Laboratory (LANL). The architecture is the result of three intersecting drivers. First, there is a general research interest in repository interoperability as exemplified by the Team's involvement in standardization efforts such as the ANSI/NISO Z39.88-2004 OpenURL Framework for Context-Sensitive Services (OpenURL) [35], the Open Archives Initiative Protocol for Metadata Harvesting (OAI-PMH) [23, 24], and more recently the Open Archives Initiative Object Re-Use & Exchange effort (OAI-ORE) [46]. Second, there is the Team's research interest in digital preservation matters illustrated by its involvement in National Digital Infrastructure and Preservation Program (NDIIPP) projects. Third, there is the concrete need to design and implement a solution for ingesting, storing and accessing the vast and growing scholarly digital collection of the Research Library of the Los Alamos National Laboratory. This paper also describes two quite distinctive implementations of the aDORe federation architecture illustrating its applicability in a variety of settings including:

- An environment operated by a sole custodian with a need to ingest, store, and access a large collection of digital objects, and where the size of the collection makes parallelized and distributed approaches a necessity.
- An environment operated by a variety of custodians, each operating their own software and hardware infrastructure but sharing a need for unified access to the union of their collections.

The remainder of this paper is structured as follows. Section 2 summarizes the results of the aDORe effort to date, and puts this paper in the perspective of previous aDORe-related communications. Section 3 describes the details of the aDORe federation approach, introducing its 3-Tier architecture, detailing the core requirements imposed on a repository to become part of a federation, and introducing the components that facilitate exposing an environment consisting of multiple, possibly heterogeneous, repositories as a single one. Section 4 is dedicated to the aDORe Archive developed and implemented at LANL in response to the aforementioned challenge to handle the Library's collection. Section 5 discusses the Ghent University Image Repository federation that is under development as a solution to the challenges posed by a large-scale, distributed, university-wide digitization effort. Both these sections describe the respective use case and how the concrete technological choices made in the deployment of the described federations relate to the high-level aDORe federation architecture. Section 6 reflects on the different implementation choices that were made in both use cases, and Section 7 concludes the paper.

## Background

The aDORe effort started at the LANL Research Library around 2003 when it became clear that the new information discovery solution for the digital library collection suffered from three significant design problems. First, the approach was metadata-centric, treating descriptive metadata records as first class citizens and the actual digital assets as auxiliary items. Second, tens of millions of digital assets were directly stored as files in a file system, resulting in a system administrator's nightmare regarding file system management and backup. Third, there was a tight integration between the content collection and the discovery application, preventing other applications from leveraging the rich content base. The solutions to these problems were straightforward and not necessarily novel: introduce a compound object view of digital assets to replace the metadata-centric view, bundle assets into storage containers that dramatically reduce the amount of files in file systems, and cleanly separate the repository from applications that leverage content hosted by the repository by providing the necessary machine interfaces. Nevertheless, the concrete implementation of these three high-level solutions led to a multi-year exploration by the Digital Library Research & Prototyping Team into the realm of repository and federation architectures. The major, self-imposed constraints throughout this effort have been to leverage existing standards and technologies to make deployment and adoption more straightforward, and to think in a distributed, component-based manner as a means to meet challenges of scale.

One strand of exploration was concerned with the choice of a compound object model and associated serialization. This led to direct involvement in the MPEG-21 standardization effort, in particular in the parts Digital Item Declaration [10], Digital Item Declaration Language and Digital Item Identification [11], a suite of papers describing the thinking with this regard [2, 6], and the release of the DIDLTools, a Java toolkit for manipulating serializations of compound objects compliant with the MPEG-21 DID data model [31].

Another strand of research investigated existing repository solutions such as Fedora [25], DSpace [37], and commercial content management systems such as XML databases. None of the investigated solutions provided adequate guarantees at the scale required by LANL. Nevertheless, architectural concepts from the Fedora effort inspired the aDORe research, and led to a regular exchange of ideas from which both efforts benefited. This exploration of repository solutions led to the XMLtape/ARCfile storage solution [29] and involvement in the WARC file [20] standardization effort.

Yet another strand of research was concerned with the nature and number of machine interfaces that are required to access materials from a repository. The distributed modeling approach automatically led to a choice of protocol-based machine interfaces and in this realm the OAI-PMH and OpenURL were leveraged [3, 4, 5, 42].

The concrete situation at LANL required a large number of XMLtapes and ARCfiles to store the collection, and naturally led to explorations in the realm of designing and implementing repository federations that expose a "single repository behavior". This federation strand is to an extent described in [3, 14, 43] but this paper provides the first overview of the aDORe federation concepts in a manner that is disconnected from specific technological choices made in the course of developing the aDORe Archive solution.

Finally, the aDORe work led to the concept of dynamically associating disseminations with stored bitstreams [3, 43]. These dynamic disseminations are the result of applying a service to a stored bitstream, and the decision regarding which services can be applied to which stored bitstreams. These decisions are guided by an on-the-fly introspection of the properties of the bitstream and of its containing compound object. This dynamic approach was dictated by considerations of scale, as the static binding of bitstreams and services (behaviors) as was proposed by the Fedora architecture led to a major maintenance overhead whenever a certain service that was statically bound to a large number of objects had to be updated.

## The aDORe Federation Architecture: Introduction

The goal of the aDORe federation architecture is to facilitate a uniform manner for client applications to **discover** and **access** content objects available in a group of distributed repositories. This is achieved by means of a 3-Tier architecture illustrated in Figure 1. Tier-3 provides client applications with a single point of access to all content available in the federation, irrespective of the actual location of that content in federated repositories. In order to realize this, the architecture requires all federated repositories to implement the same, minimal set of machine interfaces to make their content accessible. These repository interfaces constitute Tier-1 of the architecture. Moreover, the architecture requires the introduction of a middle Tier, Tier-2, consisting of two shared infrastructure components that keep the books on content objects, repositories, and repository interfaces in the federation. These shared infrastructure components minimally expose one machine interface each. In order to respond to client requests, the federation's single point of access interacts with these interfaces as well as with the interfaces exposed by the content repositories. As a matter of fact, the single point of access to the federation supports exactly the same minimal set of machine interfaces as each federated repository does, effectively making the entire federation behave in the same manner as each individual constituent repository. In principle, this design allows the aDORe federation concepts to be applied recursively, but no experiments have been conducted to date that demonstrate the feasibility of the nested federations idea. The aDORe federation architecture is not concerned with uniform operations to write, update and delete objects in repositories, and considers these the responsibility of constituent repositories of the federation. However, the architecture does ensure that results of these operations can be made apparent to client applications.

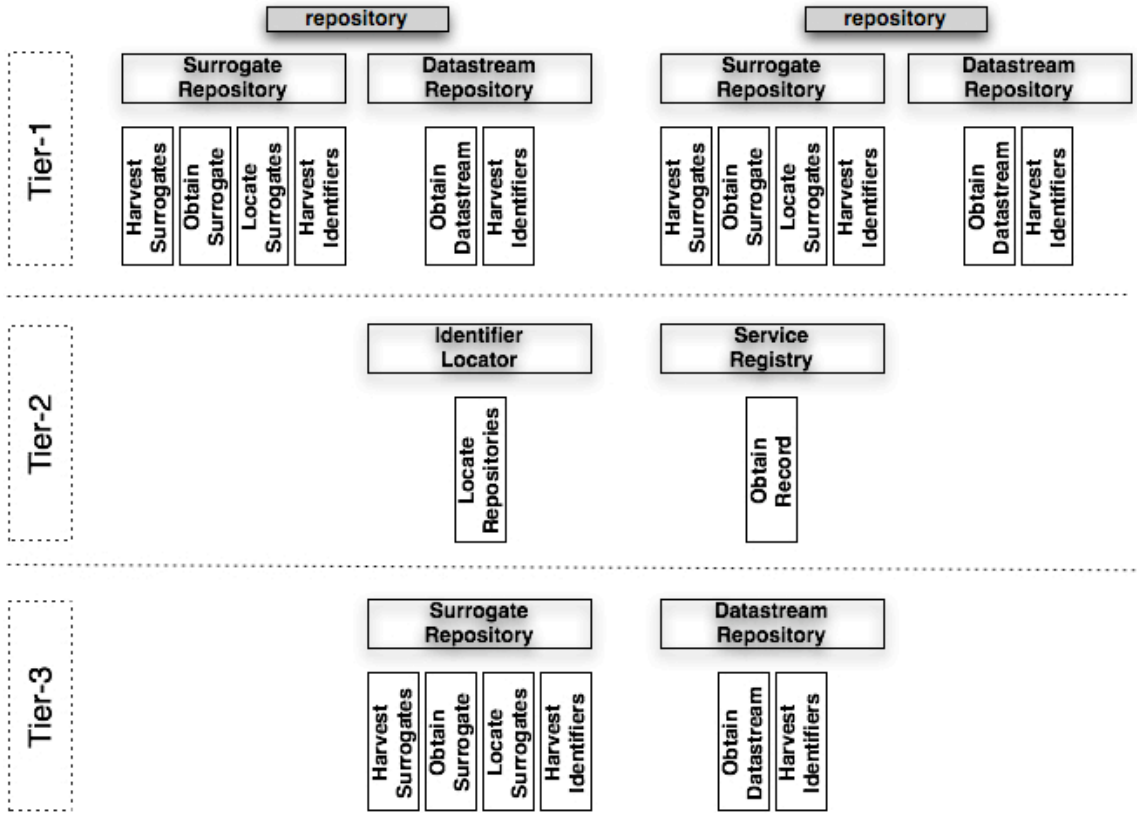

Figure 1: The 3-Tier aDORe federation architecture

# The aDORe Federation Architecture: Basic Design Choices

All entities in the aDORe federation architecture, content objects, repositories, and machine interfaces, are identified by means of URIs. The choice for URIs turns each entity into a uniquely identified resource on the Web. And an appropriate choice of the authority component of a URI scheme helps to avoid unwanted collapses of identifiers, for example, for different content objects from various federated repositories. The architecture distinguishes between protocol-based URIs that can be de-referenced via a common protocol to provide access to a representation, and non-protocol-based URIs for which no common de-referencing mechanism approach exists. The choice between these two types of URIs in the deployment of an aDORe federation relates to the use case at hand and will be explored throughout the paper.

All machine interfaces in the aDORe federation architecture are protocol-based. This choice simultaneously accommodates a multiple-custodian use case with constituent repositories that are effectively distributed across the Internet, and a single-custodian use case in which considerations of scale eventually require the distribution of components across an intranet. Although the functionality provided by the proposed machine interfaces can be implemented in a variety of ways, the desire to leverage existing standards in the aDORe work has led to using community standards that fit the job. It fact, a combination of the OAI-PMH and OpenURL can address all core requirements, and is used in both implementations of the aDORe federation architecture described below.

## The aDORe Federation Architecture: Content Objects

The architecture recognizes three types of **Content Objects**: **Digital Objects**, **Datastreams** and **Surrogates**. Certain properties related to identification, location and time-stamping of Content Objects are core enablers of the architecture, and play a crucial role in the federation's machine interfaces. Both the types of Content Objects and their core properties are described in the remainder of this section; their position in the overall architecture is also illustrated in Figure 2. It must be emphasized that the aDORe architecture does not require federated repositories to natively embrace these constructs, but rather requires supporting them in their federation-facing machine interfaces. Also, as will be shown, depending on the requirements of a specific instantiation of an aDORe federation, even some of the core properties need not be supported. The architecture supports expressing a variety of other properties and relationships pertaining to Content Objects but only serves to convey them. There is no requirement for such properties or relationships to exist, nor are any interoperability requirements imposed on them; their interpretation is left to applications overlaying the federation.

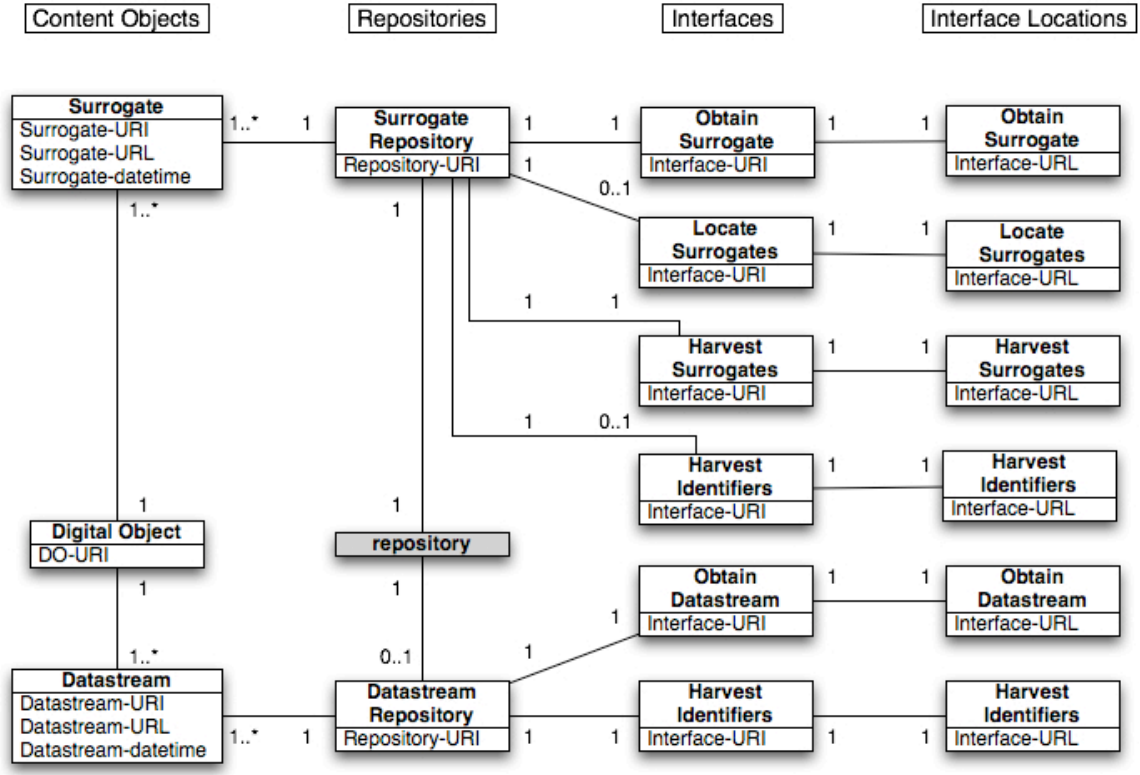

Figure 2: An overview of Tier-1 of the architecture showing the types of Content Objects, the Surrogate Repository and the Datastream Repository, as well as their core Interfaces

## Digital Objects

Compound digital objects, as initially proposed by Kahn-Wilensky [17, 18], have become the norm in digital library environments [34], and most repository systems now have some compound object model at their core. Logically, an aDORe federation also embraces compound objects, and it does so by supporting a Digital Object which is an identified aggregation of one or more Datastreams and properties pertaining to the Datastreams and to the aggregation itself. A Digital Object is the perspective of a repository's native compound digital object that is shared with an aDORe federation.

**Identification**: A Digital Object must be identified by means of a URI, the **DO-URI**. A Digital Object may have one or more DO-URIs. The DO-URI can be minted by a repository or can be inherited from another environment. Hence, a Digital Object with the same DO-URI may exist in multiple repositories of a federation. A DO-URI can be protocol-based or non-protocol-based, but in the former case the DO-URI is also *treated* as a non-protocol-based URI. This means that, in the federation environment, a DO-URI is never resolved using its native resolution protocol, but rather is conveyed as a parameter in a protocol request issued against the federation's machine interfaces. This accommodates a use case like the Internet Archive's, in which Web documents are identified in the repository by means of their native HTTP URI and where dissemination requests carry these HTTP URIs as a parameter. Example DO-URIs are *info:some-repo/do/1234* and *http://some.repo.org/do/1234*. Both are treated as non-protocol-based in a federation.

**Time-stamping**: Digital Objects can change over time, and changes are communicated to the federation by means of Surrogates and their Surrogate-datetime property.

## Datastreams

A Datastream is a retrievable bitstream of whichever media type made available by a repository to the federation. It is a perspective of a repository's native bitstream that is shared with an aDORe federation. Depending on the internal design and capabilities of a federated repository, a Datastream (retrievable bitstream) can be a straight dissemination of a bitstream stored by the repository, the dissemination of a bitstream stored external to the repository (but that the repository treats as part of the content collection it makes accessible), or the result of applying some service operation to either of those types of bitstreams. A specific Datastream can be a constituent of multiple Digital Objects made accessible by the federation, but there is only one repository in the federation from which a bitstream corresponding with the Datastream can be retrieved (i.e. there is a repository that "owns" and "serves" the Datastream).

**Identification:** A repository mints identifiers to be uniquely associated with the bitstreams it makes retrievable. These identifiers can be:

- **Datastream-URI**: A non-protocol-based URI that identifies the Datastream. Retrieval of the bitstream is achieved by using the Datastream-URI as a parameter against the appropriate machine interfaces of the federation. An example Datastream-URI is *info:some-repo/ds/5678*.

- **Datastream-URL**: A protocol-based URI that identifies the Datastream. Retrieval of the bitstream is achieved by de-referencing the Datastream-URL using its native resolution protocol. An example Datastream-URL is *http://some.repo.org/ds/5678*.

**Time-stamping**: The **Datastream-datetime** is a date/time when a Datastream underwent changes of a nature that need to be communicated to the federation. Depending on a repository, a Datastream-datetime could, for example, correspond with the time a bitstream was ingested into the repository, the time of last modification of a bitstream as recorded by a repository's file system, the time a service-operation was associated with a stored bitstream or when that service-operation was updated.

**Update policies**: Two repository policies exist that bear relationship with the Datastream-datetime:

- **New Datastream Policy**: An update of a retrievable bitstream that corresponds with a Datastream results in the introduction of a new Datastream, with a new Datastream-URI (and/or Datastream-URL) and a new Datastream-datetime. The original Datastream remains available. Under this policy, the Datastream-datetime is always the date/time of creation of the Datastream. This is a typical digital preservation scenario, in which the migration of a JPEG image identified by URI-1 results in a JPEG-2000 image identified by URI-2, not URI-1.
- **Update Datastream Policy**: An update of a retrievable bitstream that correspond with a Datastream remains associated with that same Datastream; the Datatstream-URI (and/or Datastream-URL) remains the same, but the Datastream-datetime is updated. The retrievable bitstream that originally corresponded with the Datastream is no longer retrievable. Under this policy, the Datastream-datetime is either the date/time of creation of the Datastream or the date/time of most recent modification.

## Surrogates

A Surrogate is the serialization of a Digital Object into a machine-readable representation that is made accessible by a repository. Surrogates are the vehicles repositories use to keep the federation informed about the availability of their Digital Objects and about changes those Digital Objects undergo. A Surrogate minimally expresses the DO-URI of the Digital Object of which the Surrogate is a serialization, the identifiers of constituent Datastreams of that Digital Object, as well as its own identifier. One or more Surrogates can correspond with a given Digital Object in a federation, both because a Digital Object with a specific DO-URI can exist in multiple repositories of the federation, and, because a given repository may make multiple Surrogates available for a specific Digital Object. The aDORe federation architecture allows for a choice of serialization formats such as DIDL [6, 10], METS [33], or ORE Atom [26]. Use of the same format across a federation is handy yet not essential. Still, it must be understood that a multiple format environment will impose a conversion burden either on downstream applications or on the Tier-3 components, and that format crosswalks typically lead to information loss.

**Identification:** A repository mints identifiers to be uniquely associated with the Surrogates it makes retrievable. These identifiers can be:

- **Surrogate-URI**: A Surrogate-URI is a non-protocol-based URI that identifies the Surrogate. Using a Surrogate-URI as a parameter in a protocol requests against the appropriate machine interfaces in the federation retrieves the corresponding serialization of a Digital Object. An example Surrogate-URI is *info:some-repo/su/9012*.

- **Surrogate-URL**: A Surrogate-URL is a protocol-based URI that identifies the Surrogate. Retrieval of the Surrogate is achieved by de-referencing the Surrogate-URL using its native resolution protocol. An example Surrogate-URL is *http://some.repo.org/su/9012*.

**Time-stamping:** The **Surrogate-datetime** is a date/time when a Digital Object underwent changes of a nature that needs to be communicated to the federation. Minimally, a Surrogate-datetime changes when changes the Digital Object's constituency changes, i.e. when Datastreams are added or removed. But, for those federations that implement the Datastream-URL or Datastream-datetime properties, a change to their values likely needs to be communicated, and hence will result in an update of the Surrogate-datetime. Some federations may even require an update of the Surrogate-datetime whenever any property or relationship pertaining to a Digital Object or its constituent Datastreams changes.

**Update policies**: Two repository policies exist that bear relationship with the Surrogate-datetime:

- **New Surrogate Policy**: A change to a Digital Object that needs to be communicated to the federation leads to the introduction of a new Surrogate for the Digital Object, with a new Surrogate-URI (and/or Surrogate-URL), and a new Surrogate-datetime. The previous Surrogate remains available.
- **Update Surrogate Policy**: A change to a Digital Object that needs to be communicated to the federation leads to updating the existing Surrogate for the Digital Object. The Surrogate-URI (and/or Surrogate-URL), is maintained, but its Surrogate-datetime is updated. The previous Surrogate is no longer available.

## The aDORe Federation Architecture: Tier-1

Tier-1 of the architecture, illustrated in Figure 2, consists of machine interfaces for federated repositories that support the Surrogate and Datastream notions introduced in the above, and that leverage their core properties related to identification, location and time-stamping. It should be noted that additional interfaces that leverage other properties of content objects can be added as required, but these are beyond the scope of the minimalist federation approach proposed here. In Tier-1 of the architecture, each repository exposes itself to the federation as two logical Repositories:

- A **Surrogate Repository** to facilitate access to Surrogates.
- A **Datastream Repository** to facilitate access to Datastreams.

Both types of Repositories are identified by means of a URI, the **Repository-URI**. The Repository-URI is a non-protocol-based URI that serves as a key to associate a Repository with its machine **Interfaces**. The proposed core Interfaces are discussed below, and are further illustrated in Figure 3. Each Interface is itself identified by means of a non-protocol-based URI, the **Interface-URI**, which uniquely corresponds with the network location of an Interface, the **Interface-URL**. The choice for non-protocol-based URIs to identify Repositories and Interfaces yields a stable identification across the federation, even when the network location of Interfaces changes.

As will be shown in the sections describing implementations of the architecture, Datastream Repositories are necessary when only Datastream-URIs are associated with Datastreams made available by a repository. If Datastream-URLs exist, they can directly be de-referenced using the Internet infrastructure.

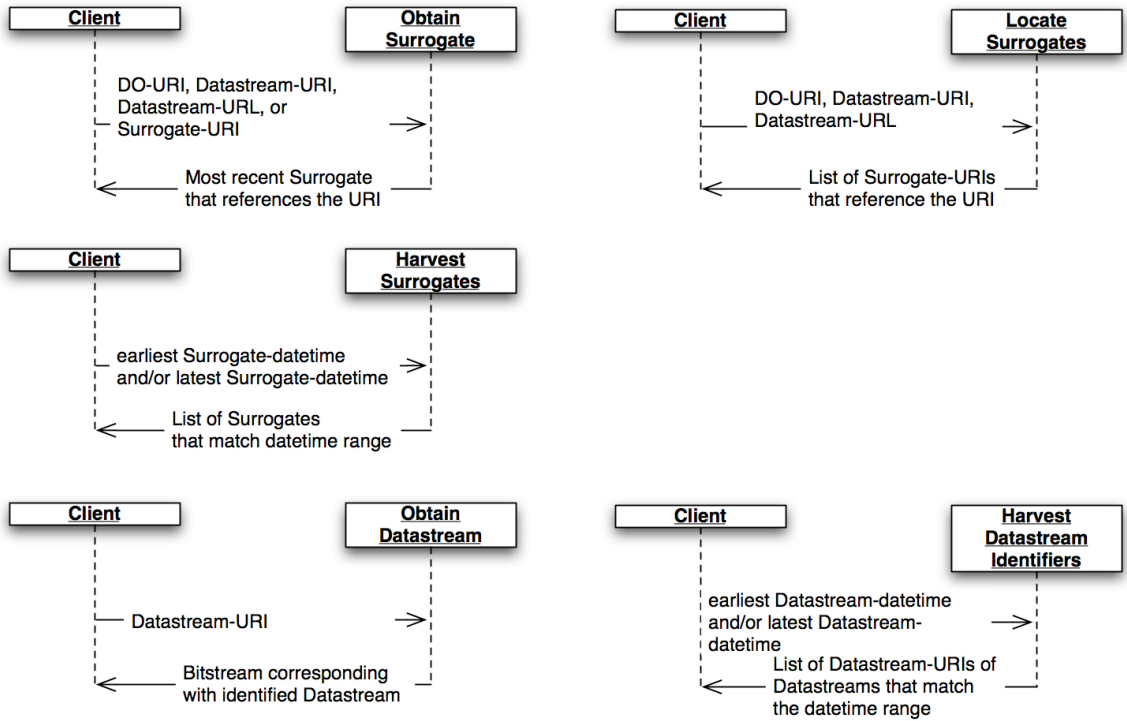

Figure 3: Core Interfaces for Surrogate and Datastream Repositories

## Surrogate Repositories: Core Machine Interfaces

Surrogate Repositories are essential for a repository to communicate the availability of Digital Objects, as well as changes applied to these Digital Objects to the federation. The proposed interfaces for a Surrogate Repository are described here.

*Harvest Surrogates*

The **Harvest Surrogates Interface** provides an essential mechanism for the federation to remain aware of Digital Objects that are available from a repository, as well as of changes in their configuration. The simplest instantiation of this Interface would return all Surrogates available from a repository in response to every request. While such an approach is possible, it seems that leveraging the Surrogate-datetime property in this Interface yields increased scalability and flexibility. Hence, the following is proposed for this Interface:

- Request parameters:
  - **from** indicating that only Surrogates with a Surrogate-datetime later than or equal to the specified date/time should be returned;
  - **until** indicating that only Surrogates with a Surrogate-datetime earlier than or equal to the specified date/time should be returned;
- Response: List of Surrogates with a Surrogate-datetime that match the specified request parameters.
- Typical implementation: OAI-PMH ListRecords with the federation's chosen Surrogate format as Metadata Format, and with Surrogate-URIs as OAI-PMH item identifiers.
  - A sample harvesting request using OAI-PMH is *http://some.repo.org/sur/oaipmh?verb=ListRecords&metadataPrefix=didl &from=2006-09-07* where *didl* indicates the Surrogate Format used in the federation.

*Obtain Surrogate*

The **Obtain Surrogate Interface** serves the purpose of obtaining a Surrogate with the most recent Surrogate-datetime that corresponds with a specified Digital Object, or with a Digital Object of which a specified Datastream is a constituent. In case Surrogates are identified by means of a Surrogate-URI, and not a Surrogate-URL, this Interface can also be used to return a Surrogate with a specified Surrogate-URI. The following is proposed for this Interface:

- Request Parameters:

- o **identifier** with a value of DO-URI, Datastream-URI, or Surrogate-URI
- Response: The Surrogate with the most recent Surrogate-datetime that corresponds with the Digital Object identified by the specified DO-URI, or that corresponds with the Digital Object of which the Datastream specified by Datastream-URI is a constituent.
- Typical implementation: OpenURL, with Referent Identifier set to DO-URI, Datastream-URI, or Surrogate-URI and with a ServiceType Identifier expressing an "Obtain Surrogate" service.
  - o A sample request using OpenURL is
    *http://some.repo.org/openurl?url_ver=z39.88-2004&rft_id=info:some-repo/do/1234&svc_id=info:ourfederation/svc/ObtainSurrogate.DIDL*
    where *ObtainSurrogate.DIDL* indicates that a Surrogate expressed using DIDL as a Surrogate Format is requested.

## Locate Surrogates

The **Locate Surrogates Interface** is relevant for repositories that have multiple Surrogates for a given Digital Object, or that have Digital Objects that share Datastreams. The Interface facilitates locating all Surrogates that correspond with a specific Digital Object, or with Digital Objects that have a specific Datastream as their constituent. The following is proposed for this Interface:

- Request Parameters:
  - o **identifier** with a value of DO-URI, Datastream-URI, or Datastream-URL
- Response: A list of Surrogate-URIs and/or Surrogate-URLs each of which identifies a Surrogate that corresponds with the Digital Object with the specified DO-URI, or with a Digital Object that has a Datastream with the specified Datastream-URI as its constituent.
- Typical implementation: OpenURL, with Referent Identifier set to DO-URI, or Datastream-URI, and with ServiceType Identifier expressing an "Locate Surrogates" service.

     o   A sample request using OpenURL is

         *http://some.repo.org/openurl?url_ver=z39.88-2004&rft_id=http://*

         *some.repo.org/ds/5678&svc_id=info:ourfederation/svc/LocateSurrogates.*

## Datastream Repositories: Core Machine Interfaces

Datastream Repositories are essential for repositories that only assign Datastream-URIs (no Datastream-URLs) to the Datastreams they make available to the federation. Using the Harvest Surrogate Interfaces of the federation will lead to discovering the existence of such Datastreams, but since the Datastream-URIs are non-protocol-based, additional information is required to de-reference them. The core Datastream Interfaces make such information available to the federation. The proposed interfaces for a Datastream Repository are described below.

### *Obtain Datastream*

The **Obtain Datastream Interface** serves the purpose of retrieving the bitstream that corresponds with a Datastream with a given Datastream-URI. The following is proposed for this Interface:

- Request Parameters:
  - o **identifier** with a value of a Datastream-URI
- Response: The bitstream that corresponds with a Datastream with the specified Datastream-URI
- Typical implementation: OpenURL, with Referent Identifier set to Datastream-URI and with a ServiceType Identifier expressing an "Obtain Datastream" service.
  - o A sample request using OpenURL is

    *http://some.repo.org/openurl?url_ver=z39.88-2004&rft_id=info:some-repo/ds/5678&svc_id=info:ourfederation/svc/ObtainDatastream.*

### *Harvest Datastream Identifiers*

The **Harvest Datastream Identifiers Interface** provides a mechanism for the federation to keep track of which Datastream-URIs are in use by the Datastream Repository (i.e.

which Datastream-URIs can be used against the Datastream Repositories' Obtain Datastream Interface). This information is used to populate the Identifier Locator of Tier-2 of the architecture. As a result, the Identifier Locator will facilitate determining to which Datastream Repository a given Datastream-URI can be submitted as a parameter. This Interface has characteristics similar to those of the Harvest Identifiers Interface of Surrogate Repositories as described above. It could be implemented in a manner whereby each request always returns all Datastream-URIs, or in a manner that allows incremental gathering of Datastream-URIs. In the latter case, the following Interface is proposed:

- Request parameters:
    - o **from** indicating that only Datastream-URIs of Datastreams with a Datastream-datetime later than or equal to the specified date/time should be returned;
    - o **until** indicating that only Datastream-URIs of Datastreams with a Datastream-datetime earlier than or equal to the specified date/time should be returned;
- Response: List of Datastream-URIs that match the specified request parameters.
- Typical implementation: OAI-PMH ListIdentifiers with Datastream-URIs as OAI-PMH item identifiers, and a Metadata Format that only expresses the Datastream-datetime. This metadata will never be requested via an OAI-PMH ListRecords request, but its choice guarantees that the OAI-PMH datestamp changes whenever the Datastream-datetime changes.
    - o A sample harvesting request using OAI-PMH is *http://some.repo.org/ds/oaipmh?verb=ListIdentifiers&metadataPrefix=datetime&from=2006-09-07* where *datetime* indicates a Metadata Format used in the federation to expresses Datastream-datetimes only.

## The aDORe Federation Architecture: Tier-2

Two shared infrastructure components, the **Identifier Locator** and the **Service Registry**, are introduced in Tier-2 of the aDORe federation architecture to manage the state of the environment, and to facilitate exposing the entire federation as a Surrogate and Datastream Repository in Tier-3.

## Identifier Locator

In its simplest instantiation, the content maintained by the Identifier Locator is a straightforward look-up table that stores the correspondence between identifiers of Content Objects available to the federation and identifiers of Surrogate Repositories and Datastream Repositories in the federation that make Content Objects with those identifiers accessible. Necessarily, the Identifier Locator will maintain this correspondence for all non-protocol-based identifiers used in the federation, as this information is essential to enable using these URIs in the Interfaces of Tier-3 of the Architecture, since Tier-3 Interfaces are not aware of either the identity of Repositories of the federation or about the network location of their Interfaces. Hence, maintained identifiers minimally include the DO-URIs, which are all treated as non-protocol-based URIs, but depending on the implementation of the architecture can also include Surrogate-URI and/or Datastream-URI. The content of the Identifier Locator is maintained by recurrently interacting with the Harvest Surrogates and Harvest Datastream Identifiers Interfaces of the federation's Surrogate and Datastream Repositories, respectively. The Identifier Locator knows about the existence of these Repositories and their Interfaces by interacting with the Service Registry.

### *Locate Repositories*

The Identifier Locator is identified by a non-protocol-based URI the **IdentifierLocator-URI**, and minimally exposes the **Locate Repositories Interface**, itself identified by means of a non-protocol-based Interface-URI with a corresponding network location, the Interface-URL. This Interface bears resemblance with the Locate Surrogates Interface described above, and hence the following is proposed:
- Request Parameters:
  - **identifier** with a value of DO-URI, Surrogate-URI, or Datastream-URI
- Response: A list of Repository-URIs of Surrogate and/or Datastream Repositories that make the Content Object with the specified identifier available.
- Typical implementation: OpenURL, with Referent Identifier set to DO-URI, Surrogate-URI, or Datastream-URI, and with ServiceType Identifier expressing an "Locate Repositories" service.

- o A sample request using OpenURL is
  *http://idlocator.ourfederation.org/openurl?url_ver=z39.88-
  2004&rft_id=http://some.repo.org/do/1234&svc_id=info:ourfederation/sv
  c/LocateRepositories.*

## Service Registry

The Service Registry keeps track of all components of the federation, as well as of their respective Interfaces. These components are all Surrogate and Datastream Repositories of the federation, and also the Identifier Locator, the Service Registry itself, and the Repositories introduced in Tier-3 of the architecture. In essence, the content consists of two lookup tables, one listing the correspondence between the URI of a component (e.g. Repository-URI) and its matching Interface-URIs, the other listing the correspondence between these Interface-URIs and their Interface-URLs. Note that the type of Interface is expressed in the first look-up table, in order to allow client-applications (typically the components of Tier-3 or the Identifier Locator) to select the appropriate Interface for the task at hand.

### *Obtain Registry Record*

The Service Registry is identified by a non-protocol-based URI the **ServiceRegistry-URI**, and minimally exposes the **Obtain Registry Record Interface**, itself identified by means of a non-protocol-based Interface-URI with a corresponding network location, the Interface-URL. The following is proposed for this Interface:

- Request Parameters:
  - o **identifier** with a value of the URI of a component (e.g. Repository-URI), or of an Interface-URI.
- Response: A list of Interface-URIs and corresponding Interface-type that match the specified component URI, or the Interface-URL that corresponds with the specified Interface-URI.
- Typical implementation: OpenURL, with Referent Identifier set to the URI of the component or of the Interface, and with ServiceType Identifier expressing an "Obtain Registry Record" service.

- o A sample request using OpenURL is
  *http://svcregistry.ourfederation.org/openurl?url_ver=z39.88-*
  *2004&rft_id=info:some-repo/*
  *&svc_id=info:ourfederation/svc/ObtainRecord.*

## The aDORe Federation Architecture: Tier-3

In Tier-3, the entire federation is presented to downstream applications as a single Surrogate Repository, and, depending on the implementation, an additional single Datastream Repository. These Repositories have exactly the same Interfaces as described in Tier-1. Applications overlaying the federation only need to know about the existence of the federation's single Surrogate and Datastream Repository to build upon the content made available in all federated repositories that are effectively hidden from them.

The Surrogate and Datastream Repositories of Tier-3 can support the core Surrogate and datastream Interfaces, respectively, by interacting with the appropriate Interfaces of Tier-2 components and Tier-1 Repositories. For example, presume an overlay client uses the Locate Surrogate Interface of the Tier-3 Surrogate Repository in order to find all Surrogates in the federation that correspond with a specific DO-URI. In order to generate a response, the Tier-3 Surrogate Repository first issues a request against the Identifier Locator's Locate Repositories Interface with this DO-URI as parameter, and receives a list of Repository-URIs of Tier-1 Surrogate Repositories that expose Surrogates for the given DO-URI in response. Next, for each of these Repository-URIs, the Tier-3 Surrogate Repository does a look-up in the Service Registry to find the network location of the Locate Surrogate Interface for the identified Repository. At this point, the Tier-3 Surrogate Repository can respond to the client with a list of Locate Surrogate requests each carrying the DO-URI as a parameter and targeted at a Tier-1 Surrogate Repository that was listed in the response from the Identifier Locator. The client can now issue each requests itself, and build a list of all matching Surrogates in the federation understanding that a single Surrogate Repository may expose multiple Surrogates for a given Digital Object.

Alternatively, the Tier-3 Surrogate Repository could issue all these requests, merge all responses and return the resulting list to the client. Whichever approach is taken, the client can now retrieve all Surrogates corresponding with the specified DO-URI. In an environment where Surrogate-URIs are used, this is achieved by using these URIs as a parameter in requests against the Tier-3 Surrogate Repositories' Obtain Surrogate Interface. If Surrogate-URLs are used, they can be de-referenced using the Internet infrastructure.

# The aDORe Archive

## Use Case

The Research Library of the Los Alamos National Laboratory (LANL) hosts a significant digital scholarly collection and makes services based on that collection available to its customer base. The collection currently consists of licensed content from both secondary and primary publishers (e.g. APS, BIOSIS, EI, Elsevier, Thomson Scientific, etc.) and unclassified LANL Technical Reports, and is expected to grow to include a wide variety of unclassified digital assets that result from the Laboratory's research endeavors. As explained in the Background Section, previous incarnations of the Library's repository had fallen victim to issues of scalability. A uniform approach for ingesting, storing, and disseminating content was necessary to ensure the collection's manageability, accessibility, and preservation.  Also, the sheer volume of the collection required parallelization for ingestion and dissemination, and distribution for storage.

The aDORe Archive was designed and developed in response to this challenge. It is a major source of inspiration for high-level federation concepts described above. The aDORe Archive software is available for download from the aDORe project site [30], and illustrates the benefit of consistently using standards throughout a software solution, as doing so allows the re-use of major building blocks developed by third parties.  For example, OCLC's OAI-PMH and OpenURL packages have been used throughout the aDORe Archive solution. The remainder of this section categorizes the aDORe Archive in terms of the aDORe federation concepts introduced above. Figure 4 illustrates the architectural relationship, and Table 1 and Table 2 provide a summary of choices regarding Content Objects and Interfaces, respectively.

Some core characteristics of the aDORe Archive are a direct result of its write-once/read-many approach that was motivated by the batch manner in which LANL typically obtains content from publishers. Interestingly enough, those characteristics are also appealing for digital preservation scenarios. The fundamental storage components in the aDORe Archive are ARCfiles and XMLtapes. ARCfiles were introduced by the Internet Archive as a means to concatenate large amounts of documents resulting from a Web crawl into a single file (the ARCfile). Individual documents are made accessible through APIs that leverage indexes external to the ARCfile. ARCfiles are used in the aDORe Archive as a container to store constituent bitstreams of Digital Objects. XMLtapes are similar to ARCfiles, but are well-formed XML files that concatenate large amounts of Surrogates. As is the case with ARCfiles, documents in XMLtapes can be accessed via APIs and indexes external to the XMLtapes. Since XMLtapes are XML files, they can also be handled using off-the-shelf XML tools. Both ARCfiles and XMLtapes are read-only storage components.

When ingesting a batch of compound objects, an XML-based Surrogate corresponding with each object is created, and the resulting Surrogates are concatenated into one or more XMLtapes. Similarly, the bitstreams of the batch of compound objects are concatenated into ARCfiles. It is worthwhile to note the handling of different configurations of a same Digital Object. Examples of such different configurations include different (publication) versions of a Digital Object that share a DO-URI, and different Premis representations [7, 27] of a same Digital Object. These Premis representations vary in their constituent Datastreams as a result of the migration of some underlying bitstreams and the introduction of a new Datastream for such migrated bitstreams. Ingesting a new configuration of a previously ingested Digital Object is treated as any other ingestion: no checking is performed as to whether a Digital Object with a specified DO-URI already exists, and a new Surrogate with a new Surrogate-URI and new Surogate-datetime is created. Updating a Digital Object, for example, because a constituent bitstream needs to be migrated, is treated as the combination of retrieving both the most recent Surrogate for the Digital Object and the problematic bitstream, followed by ingesting a Digital Object that shares all characteristics with the initially retrieved one, with the exception of having the migrated bitstream as a constituent Datastream. The new Digital Object will have the same DO-URI(s), but will be instantiated as a new Surrogate, with a new Surrogate-URI and a new Surrogate-datetime. The various Surrogates for a given Digital Object exist autonomously in the Tier-1 repositories of the aDORe Archive, but can be joined through intermediation of Tier-2's Identifier Locator that, among others, keeps track of the location of all repositories that host a Digital Object with a specific DO-URI. Note that this approach allows dynamically constructing an audit trail of the various configurations of a Digital Object.

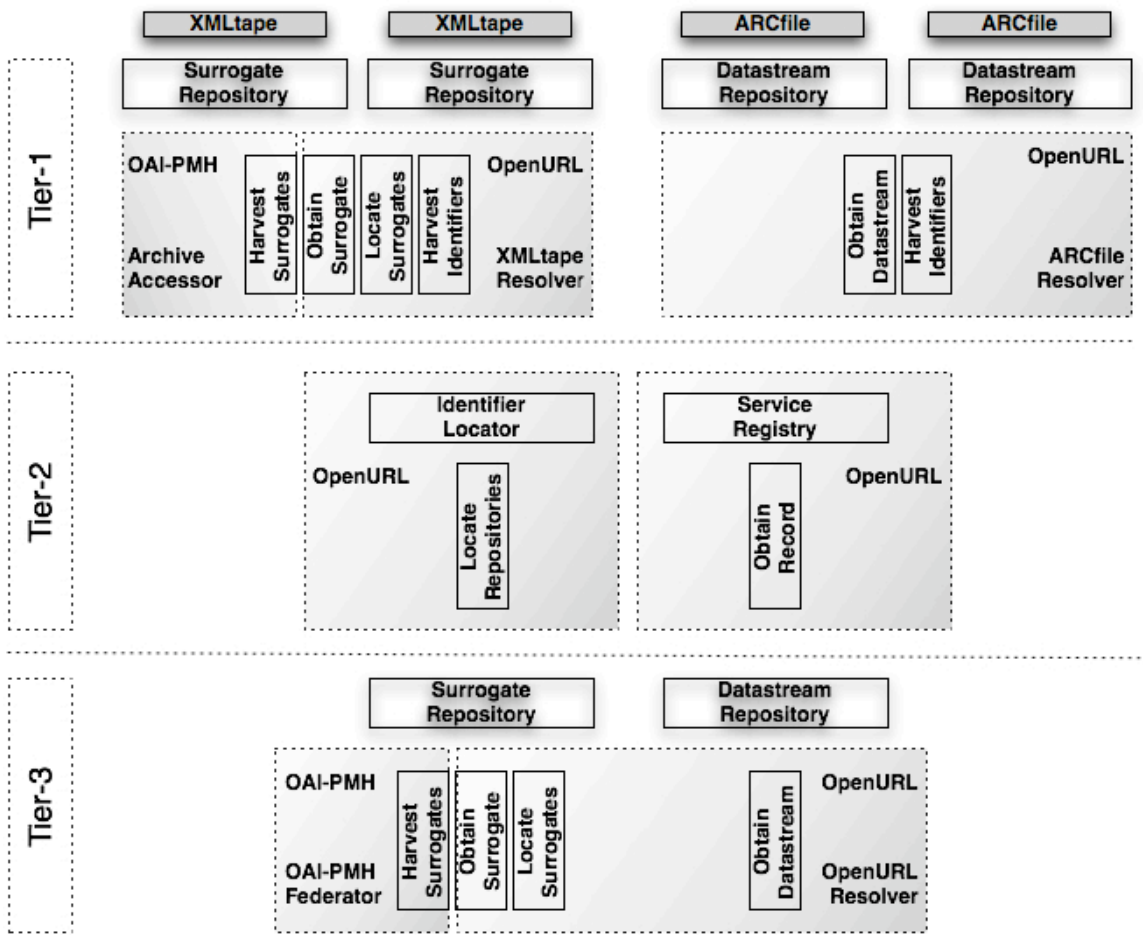

Figure 4: The aDORe Archive

## Content Objects

The Digital Objects at the LANL Research Library are scholarly artifacts (e.g. journal papers) or descriptions of these artifacts (e.g. records from abstracting and indexing databases).  In all cases, they are compound, consisting of multiple bitstreams. In order to implement a common representation approach for the Digital Objects in LANL's aDORe Archive deployment, MPEG-21 DIDL was chosen as a Surrogate Format.  It should be noted, however, that the aDORe Archive software itself is neutral regarding a choice of Surrogate Format. Datastreams of the aDORe Archive directly correspond with stored bitstreams.

At ingestion time, all Content Objects are assigned non-protocol-based URIs in the *info:lanl-repo/* namespace, resulting in an environment that achieves a complete virtualization (repositories can be moved around at will) but that requires additional components for URI de-referencing. For Surrogates and stored bitstreams, the values for these URIs as computed using the UUID algorithm [28]. For Digital Objects, the values for the *info:lanl-repo/* URIs are typically derived from the publishers' non-URI identifiers  (e.g. Inspec identifiers). In addition to that, Digital Objects inherit URIs that were assigned by publishers, such as DOIs (expressed as URIs in the *info:doi/* namespace) or HTTP URIs. Note that such URIs are always treated as non-protocol-based, even if they were minted in a protocol-based URI scheme such as HTTP. The identifiers listed by Surrogates in the aDORe Archive are DO-URIs, Surrogate-URIs, and Datastream-URIs. No Surrogate-URLs or Datastream-URLs are listed. Retrieval of Surrogates or Datastreams is achieved via the appropriate Interfaces.

The New Surrogate and New Datastream Policies of the aDORe Archive are a direct result of the write-one/read-many approach described above, but are maintained in storage approaches other than XMLtape/ARCfile that are under development for the aDORe Archive.

**Tier-1**

A typical content repository in the aDORe Archive is an XMLtape or an ARCfile. These directly correspond with a Surrogate Repository and a Datastream Repository of the aDORe federation architecture, respectively. The Interfaces for these Repositories leverage the APIs of the underlying storage components. However, other repository types can be added. For example, in order to meet the need to ingest objects one at a time, instead of in batch mode, a storage solution combining a relational database that stores Surrogates as blobs (Surogate Repository), and a file-system with appropriate directory structure that stores individual bitstreams (Datastream Repository) was recently developed. In all cases, all core Tier-1 Interfaces were implemented, hiding the underlying repository technology, and providing consistent protocol-based access to Surrogates and Datastreams irrespective of the repository type. All Repositories and Interfaces are identified by means of URIs in the info:lanl-repo/ namespace with a value generated by the UUID algorithm.

Since an aDORe Archive is designed to host a large amount of XMLtapes and ARCfiles (already in the order of 10,000 at the time of writing in the LANL deployment) a solution was devised that provides a single-point of access for each core Interface of all XMLtapes and ARCfiles, respectively, rather than a separate Interface for each. This is achieved by introducing a registry of XMLtapes and ARCfiles. In addition to the core Interfaces, the aDORe Archive also provides a generic XQuery capability that allows collection administrators to issue ad hoc queries against individual Surrogate Repositories.

**Tier-2**

In Tier-2, the Service Registry keeps track of the Repositories of Tier-1, as well as of the identity, type and location of their Interfaces. In addition to this basic information, the Service Registry also stores a variety of metadata pertaining to the collections made accessible by the Repositories. This metadata is typically associated with a batch of Digital Objects at ingestion time, and along with the Repository-URIs, Interface-URIs and Interface-URLs, it is registered into the Service Registry during the ingestion process. The Service Registry stores information in a manner that is compatible with the IESR specification [1, 16], and its implementation is based on the Ockham Registry software. It provides the core Obtain Registry Record Interface, but also supports harvesting and searching via OAI-PMH and SRU Interfaces, respectively.

Also in Tier-2, the Identifier Locator stores the correspondence between DO-URIs, Surrogate-URIs, and Datastream-URIs on one hand, and Repository-URIs on the other. It is populated by interacting with the Datastream Repositories' Harvest Datastream Identifiers Interface, and with a special-purpose Harvest Identifiers Interface that was introduced for Surrogate Repositories as an optimization to harvesting identifiers via the Harvest Surrogates Interface. For each XMLtape and ARCfile added to the environment this interaction takes place at the very end of the ingestion process. For Repositories such as the aforementioned MySQL/file-system combination, identifiers are collected on a recurrent basis. The Identifier Locator is implemented as a highly optimized instance of MySQL that provides sub-10ms responses for its Locate Repositories Interface. At the time of writing the Identifier Locator stores over 400,000,000 URIs of Content Objects.

Tier-2 of the aDORe Archive also contains Registries that standardize property vocabularies across the environment. The Format Registry lists locally assigned URIs to identify bitstream types and flavors of XML, and associated metadata including format identifiers assigned by other authorities (e.g. MIME media types and Pronom identifiers). The Semantic Registry lists locally assigned URIs used to semantically characterize Content Objects, and associated metadata that mainly consists of a human readable explanation of what the semantic URI stands for. Commonly used URIs characterize bitstreams as a full-text scholarly paper, a bibliographic description of a scholarly paper, or a reference made in a scholarly paper. Both Registries have machine interfaces based on OAI-PMH and OpenURL.

**Tier-3**

In Tier-3, aDORe Archive's front-ends are introduced to serve as sole gateways to the Tier-1 repositories: the OAI-PMH Federator implements the Harvest Surrogate Interface for the entire environment, whereas the OpenURL Resolver implements the remaining core Surrogate and Datastream Repository Interfaces. In order to respond to requests, both front-ends first interact with the Identifier Locator and Service Registry of Tier-2, and next with the Interfaces of the Repositories of Tier-1. A rule-based engine that dynamically associates service-driven disseminations with stored bitstreams powers the OpenURL Resolver. This functionality is exposed by an additional Interface that allows requesting a list of available disseminations for any URI-identified Content Object. In this list, all available disseminations are expressed as dissemination requests directed at the same Interface [5].

Table 1: Content Objects in the aDORe Archive and the Ghent Image Server Federation

| Content Object | Property | aDORe Archive | Ghent Image Server Federation |
|---|---|---|---|
| Digital Object | | | |
| | DO-URI | URIs in the info:lanl-repo/ namespace minted during ingestion, and URIs (e.g. DOIs) inherited from other environments. | URIs in the info:ugent-repo/ namespace minted during ingestion. |
| | Digital Objects with same DO-URI in federation? | Multiple publication versions and multiple Premis representations of the same object share a DO-URI. | DOs can be fragmented over multiple repositories. |
| | Digital Objects with same Datastreams in federation? | Digital Objects can share Datastreams although this is currently not the case. | DOs can in theory share Datastreams although this is currently not the case. |
| Surrogate | | | |
| | Surrogate-URI | URIs in the info:lanl-repo/ namespace minted during ingestion. | URIs in the info:ugent-repo/ namespace that leverage internal identifiers assigned by the repositories involved. |
| | Surrogate-URL | n/a | n/a |
| | Surrogate-datetime | Datetime of Surrogate creation | Datetime of most recent change to Digital Object |
| | New Surrogate Policy | A new Surrogate is created to reflect a different configuration of a Digital Object. | n/a |
| | Update Surrogate Policy | n/a | Existing Surrogate is updated to reflect a different configuration of a Digital Object. |
| | Surrogate Format | MPEG-21 DIDL | MPEG-21 DIDL |
| Datastream | | Only stored bitstreams. | Only service-based disseminations of stored bitstreams. Stored bitstreams |
| | | | not accessible. |
| | Datastream-URI | URIs in the info:lanl-repo/ namespace minted during ingestion. | n/a |
| | Datastream-URL | n/a | KEV OpenURLs with DO-URI as Referent Identifier and an indication of the requested service (e.g. GetThumbnail) as the ServiceType Identifier. |
| | Datastream-datetime | Datetime of  ingestion of bitstream. | Date/time of associating the service-based dissemination with a stored bitstream. |
| | New Datastream Policy | Yes, but not implemented in practice yet. | n/a |
| | Update Datastream Policy | n/a | Yes. |

Table 2: Interfaces in the aDORe Archive and the Ghent Image Server Federation

| Repository | Interface | aDORe Archive | Ghent Image Server Federation |
|---|---|---|---|
| Surrogate Repository | | Available for all XMLtapes. | Available for both eRez and Aleph. |
| | Harvest Surrogates | OAI-PMH with MPEG-21 DIDL as Metadata Format. | OAI-PMH with MPEG-21 DIDL as Metadata Format. |
| | Obtain Surrogate | KEV OpenURL with DO-URI, Surrogate-URI, or Datastream-URI as Referent Identifier. Response is DIDL. | KEV OpenURL with DO-URI or Surrogate-URI as Referent Identifier. Response is DIDL. |
| | Locate Surrogates | KEV OpenURL with DO-URI, Surrogate-URI, or Datastream-URI as Referent Identifier. SRU XML Response containing the URI that was used as the value of Referent Identifier and the corresponding Repository-URI. | So far, no use case has been identified that requires implementing this Interface. |
| Datastream Repository | | Available for all ARCfiles. | No Datastream Repositories. |
| | Obtain Datastream | KEV OpenURL with Datastream-URI as Referent Identifier. | n/a |
| | Harvest Datastream Identifiers | KEV OpenURL with Repository-URI as Referent Identifier. Response is a plain text list of identifiers, delimited by new line character. | n/a |

# The Ghent Image Repository Federation

## Use Case

In 2006, Ghent University started providing funds for digitizing image collections held by departments across the campus. These collections consist of a wide variety of materials including slides, maps, x-rays, hard copies of material used in university courses, and syllabi, and each holds anywhere between a few hundred to tens of thousands of objects. In digitized form, collection sizes range between a few gigabytes to several terabytes. Early estimates indicate an annual data growth of about 8 terabytes, overall. In addition to this, in 2007, the Ghent University Library signed a partnership with Google Books [40] that will result in the digitization of three hundred thousand books that eventually will be made part of the university's content network.

The results of the digitization efforts are managed in a variety of ways. Some departments remain custodians of their collections, operating them on a content management system of their choosing. Other departments lack the resources or enthusiasm for in-house management, and make use of a centrally provided storage and management facility. Still, within this hybrid environment, Ghent University aims at maximizing return on investment, and wants to avoid a fragmented landscape that prevents straightforward use of materials across departmental and software boundaries. For example, all materials must be directly accessible in the university's Minerva e-Learning environment. Hence, a solution is required that allows for consistent discovery and re-use of the outcomes of the massive digitization effort.

In response to this challenge, the Ghent University Library has embarked on a pilot project that uses aDORe federation concepts as the design guideline. Unlike the aDORe Archive case described above, in which all repositories largely share the same design (XMLtapes and ARCfiles), and are managed by the same custodian, the Ghent Library takes heterogeneity as the starting point. It works towards a solution whereby all media management systems across campus can be taken on board, and where each can continue providing its native functionality to the target customer base. However, in order to achieve a unified perspective of the distributed collection, and to allow cross-system applications, the Library's strategy is based on extending each system with core Interfaces proposed by the aDORe federation architecture, and to implement some of its Tier-2 and Tier-3 components. In the ongoing pilot, the Library incorporates two repositories: the commercially available eRez imaging server that hosts about 40,000 scanned images, a total of about 2 terabytes, and Ex Libris' Aleph catalogue system that, among others, hosts the bibliographic metadata pertaining to these images. The Picture Database application [39] overlays both repositories, and exemplifies an application that could eventually be deployed across Ghent University's distributed image management systems.

The remainder of this section categorizes the Ghent Image Repository federation in terms of the aDORe federation concepts introduced above. Table 1 and Table 2 provide a summary of choices regarding Content Objects and Interfaces, respectively.

## Content Objects

The Digital Objects in the Ghent pilot are the digitized images of the eRez server on one hand, and their bibliographic description as maintained by Aleph, on the other. The eRez server stores TIFF master images, and implements the concept of *single source dynamic imaging,* which facilitates dynamically generating image variations and common media types from a single master. As a matter of fact, the TIFF master itself is never made accessible by eRez, only its service-based transforms are. As a result, the Datastreams that eRez exposes to the federation are not the stored TIFF bitstreams but their service-based transforms. Each Datastream is only identified by means of a Datastream-URL, which is an OpenURL that contains both the eRez identifier of the TIFF and the indication of the requested service as parameters. Each TIFF master is the seed for a Digital Object that consists of a set of Datastreams, each of which is a service-based transform of the master. The amount and nature of available Datastreams for any given Digital Object is dynamically decided in a rule-engine based process inspired by the one described in [3]. The eRez server allows attaching IPTC [12] and EXIF [13] metadata to stored masters, but the Ghent Library preferred to use the existing Aleph cataloguing environment for manually generated metadata. Each Datastream for the Aleph system is a MARCXML record describing an image master and is identified by a Datastream-URL only. Digital Objects in Aleph consist of this single Datastream only. Both eRez and Aleph use the same DO-URI to identify Digital Objects that pertain to the same TIFF master, indicating that both repositories have part of the perspective on any given object, and allowing merging of perspectives in overlaying applications. The DO-URIs are expressed in the *info:ugent-repo/* namespace, and actual URIs combine an appropriate string that identifies the pilot project, and an identifier minted during the ingestion process. Both eRez and Aleph use MPEG-21 DIDL as the Surrogate Format, and both systems dynamically generate their Surrogates upon request. Surrogates are uniquely identified by means of Surrogate-URIs, again expressed using the info URI scheme, that combine a string identifying the repository that exposes the Surrogate (eRez or Aleph), and an internal identifier minted by those repositories. The Surrogates list DO-URI, Datastream-URLs, and the Surrogate-URI as identifiers. The dynamic nature of deciding on the constituent Datastreams of an eRez Digital Object, and of generating Surrogates for both eRez and Aleph yields an environment that adheres to the Update Surrogate Policy. Only Surrogates that denote the current configuration of a Digital Object are available. Also, the dynamic generation of disseminations in eRez, and the overwrite-approach of Aleph that is typical of cataloguing systems, leads to an Update Datastream Policy for both repositories.

**Tier-1**

The content repositories in the current pilot are the eRez and Aleph systems, but will eventually include the image management systems operated across Ghent University. For both eRez and Aleph, Surrogate Repositories based on OCLC's OAI-PMH package were implemented that support all proposed Surrogate Interfaces. For Aleph, the implementation was straightforward and was based on one of the many examples provided in OCLC's software that detail connecting with a relational database. For eRez, implementation was less obvious since the system has no relational database but rather a Lucene search engine as its back-end for accessing stored objects. In essence, three main requirements must be met in order to implement OAI-PMH for these types of systems:

the system must have an index for document identifiers, an index for document datestamps, and it must support a query that returns all documents. The latter requirement was the most challenging and was tackled by developing an XML-based search API that serves as the access point for OCLC's OAI-PMH package. The API leverages the datestamp indexes and specially crafted eRez templates. With this API in place, providing the OAI-PMH-based Surrogate Repository was straightforward: incoming Harvest Surrogate Requests are mapped to eRez API calls that fetch image metadata as well as URIs for all associated Datastreams (dynamic disseminations of the stored image); all resulting information is then written into MPEG-21 DIDL Surrogates that are returned to the harvesting client. Obtain Surrogate interfaces for both systems are provided by a home-grown OpenURL servlet. For eRez, a DO-URI provided on an OpenURL request is first submitted as a search term to the aforementioned XML API. The response is a Surrogate-URI that is then used by the OpenURL servlet as the key on a GetRecord request submitted to the eRez OAI-PMH repository. The resulting MPEG-21 DIDL Surrogate is returned to the client. For Aleph, an extra index had to be added to the database to resolve DO-URIs to Surrogate-URIs. Once a Surrogate-URI is available, the Aleph OAI-PMH repository is used in the same manner as described for eRez. Since all Datastream identifiers are protocol-based, no Datastream Repositories had to be introduced.

## Tier-2

The simplicity of the pilot environment and the fact that the same custodian operates both repositories as well as the overlaying Picture Database application, did not call for the introduction of a Service Registry. However, as soon as the federation will be extended to include a centrally operated eRez system to serve departments that prefer not to locally manage their image collections, this shared infrastructure component will be introduced. At that point, an Identifier Locator that supports requesting a Surrogate for any DO-URI used in the federation will also be introduced.

**Tier-3**

A harvester whose task it is to create and maintain a central cache of all Surrogates of the federation will be the initial client of the Service Registry. This central cache will be the single point of access to harvest Surrogates from the entire federation. It corresponds to the Tier-3 Surrogate Repository of the aDORe federation architecture, and will support all core Interfaces. The Identifier Locator will actually be populated by harvesting from this Tier-3 Surrogate Repository instead of from all Tier-1 Repositories as is the case in the aDORe Archive that maintains no centralized Surrogate cache but rather dynamically polls all appropriate Surrogate Repositories of the federation to respond to harvesting requests. The information stored by the Identifier Locator will allow implementing an OpenURL-based Obtain Surrogate Interface, which returns a Surrogate for any DO-URI used in the federation.

## Discussion

The major distinction between the aDORe Archive and Ghent Image Repository federation is the omission of Datastream Repositories in the latter, as a result of a choice for only protocol-based URIs to identify Datastreams. When working with repositories that are distributed across the Internet, this choice is quite sensible because the identifying Datastream-URLs can be de-referenced using the available Internet infrastructure and without additional know-how regarding a special-purpose de-referencing infrastructure that is required when Datstream-URIs are chosen to identify Datastreams. Nevertheless, in environments such as the aDORe Archive that have some long-term digital-preservation aspirations, the long-term horizon yields concerns about a tight coupling between identifier and identifier de-referencing as established by protocol-based URIs. This concern is motivated by practice that shows that access URLs for repository objects change over time as a result of technical, policy or custodianship issues. Meanwhile, the internal identification assigned to these objects remains stable even across generations of content management systems. In this case, non-protocol-based URIs that leverage the stability of those internal identifiers, but are turned into URIs of non-protocol-based schemes such as info [44], ARK [21], and tag [19] are appealing because they introduce both global uniqueness and a level of *virtualization* (i.e. identifiers of Content Objects can remain stable, while the physical location of the objects can change over time). Also, non-protocol-based URIs allow intentional collapses of identifiers. Such collapses are useful when multiple repositories hold a copy of the same object and use the same identifier for it, as can be the case in preservation scenarios. They are also of interest to cases where a single repository holds multiple copies of an object with the same identifier; the Internet Archive serves as an example. Protocol-based URIs effectively makes such wanted collapses impossible.

Another noteworthy design difference between the two cases is the introduction of a Surrogate cache in the Ghent case to implement the Tier-3 Surrogate Repository. In the aDORe Archive, no such cache is created as the Tier-3 Surrogate Repository responds to Harvesting requests by dynamically harvesting from the appropriate Tier-1 Surrogate Repositories. Again, Ghent's choice is sensible in the context of the operating environment that consists of multiple, distributed repositories with one likely being more reliable and responsive than the other. As already described in [14], the dynamic harvesting approach taken in the aDORe Archive can successfully be deployed in Intranet environments, but may cause problems in truly distributed set-ups where a harvesting session against a federation's Tier-3 Surrogate Repository may fail only because one of the federated repositories fails to respond. The larger the federation becomes, the higher the chances of such failures become, indicating a problem of scale with the federation. Ghent's approach avoids this problem through the creation of a central cache that becomes the single point of access for harvesting from the federation. An alternative is to disclose the Tier-2 Service Registry to overlaying applications, and allow those to build their own harvesting strategies, and directly harvest from Tier-1 Surrogate Repositories. This approach is especially attractive when the Service Registry has an additional search Interface and rich registry records that detail the nature of the each repositories' collection.

Another concern of scale in the federation pertains to the Identifier Locator. Indeed, the size of the database underlying the Identifier Locator depends on the amount of Content

Objects in a federation, on whether only Digital Objects are identified by means of non-protocol-based URIs or whether all Content Objects are. It also depends on whether the Identifier Locator maintains auxiliary data such as Surrogate-datetime, Datastream-datetime, or for informative purposes, even Surrogate-URLs and Datastream-URLs. The aDORe Archive example illustrates that the Identifier Locator database can grow to such an extent that eventually, in its own right, it becomes subject to distribution and federation. That is why, in the aDORe Archive, the Identifier Locator is implemented using multiple MySQL instances running on a blade server environment, and a front-end that allows querying the entire set-up. In an Internet environment, distribution of the Identifier Locator can also be achieved, for example, by having each Repository operate its own Identifier Locator. This approach removes the need to harvest identifiers into a central environment, but introduces the need for reliable approach to query across the distributed Identifier Locators. This could, for example, be achieved by means of the introduction of a distributed search application in Tier-2 of the architecture, which would effectively replace the shared Identifier Locator.

Finally, it is worth noting that the choice of Surrogate update policy is likely to influence the choice of Surrogate-URIs. Indeed, the aDORe Archive follows the New Surrogate Policy, making a different Surrogate available to correspond with the various configurations of a Digital Object. In this case, Surrogate-URIs are orthogonal to DO-URIs. The Ghent Image Server Federation follows the Update Surrogate Policy, making one Surrogate available for each Digital Object, which only reflects the most recent configuration of the Digital Object. In this case, Surrogate-URIs and DO-URIs could be chosen to coincide. A Fedora repository meticulously records an audit trail of the changes that a Fedora object undergoes. Assuming a one-to-one correspondence between a Fedora object and a Digital Object, this creates two ways in which Fedora could implement Surrogates. It can associate a single Surrogate with a Fedora object, in which case Fedora would adhere to the Update Surrogate Policy, but interestingly enough, each Surrogate would convey all configurations of the associated Digital Object. In this case, the Surrogate-URI could coincide with the DO-URI. Alternatively, Fedora can associate multiple Surrogates with a Digital Object, one per configuration, in which case Fedora would follow the New Surrogate Policy. In this case, the Surrogate-URI could be some unique combination of a DO-URI and an audit trail date/time.

## Conclusion

The starting point of this paper was the consideration that the need to federate repositories naturally occurs in two distinct environments. One is characterized by the existence of a single custodian in charge of managing a vast digital object collection in an Intranet context, the other by multiple custodians each operating a collection of interest to some community or application, with hosting repositories distributed across the Internet. This paper has detailed the core concepts of the high-level aDORe federation architecture, and has shown examples of two federations whose design and implementation was guided by the architecture. In Tier-1, repositories expose common interfaces that leverage two properties of content objects: identifiers and timestamps. By restricting interfaces to only these two core properties, the architecture imposes minimal interoperability requirements on federated repositories, but, as a result, requires cross-federation applications to address requirements that pertain to other properties. The Tier-2 components, Identifier Locator and Service Registry, actually bind the individual repositories of Tier-1 into a federation as they facilitate discovering identifiers and services across those repositories. As a matter of fact, these two tiers suffice to make a federation operational. However, in certain use cases, a "single repository behavior" may be required for the entire federation; this is achieved by introducing Tier-3. This tier removes complexity for clients of the federation, but introduces challenges especially related to harvesting Surrogates from all federated repositories via a single interface [14].

To an extent, the issues that were raised in this paper, and the solutions that were proposed may come across as of interest in only a marginal set of use cases. Interestingly enough, when taking a parochial perspective of the repository landscape they may indeed be. However, when looking at repositories from a collective perspective in which distributed repositories are regarded the basis of a future scholarly communication infrastructure [41, 45, 47], the solution to certain requirements lies in federating. For example, after approximately ten years of global institutional repository efforts, there still is no reliable and comprehensive infrastructure that allows locating a self-archived and hence freely available copy of a paper with a known Digital Object Identifier. To an extent this is due to the mistreatment of pre-existing identifiers of scholarly materials as second-class metadata upon ingestion in repositories. To a larger extent, this is due to the lack of collective, federated thinking.

# Acknowledgments

Herbert Van de Sompel acknowledges the fundamental contributions that were made to the aDORe effort by past and present members of the Digital Library Research and Prototyping Team: Lyudmila Balakireva, Jeroen Bekaert, Ryan Chute, Patrick Hochstenbach, Henry Jerez, Xiaoming Liu, and Kjell Lotigiers. Many thanks go out to our Fedora colleagues at Cornell University for inspiration and appreciation: Carl Lagoze and Sandy Payette. And many thanks also to Michael Nelson at Old Dominion University for proofreading a draft of this paper.